\documentclass[sn-mathphys,Numbered]{sn-jnl}

\usepackage{graphicx}%
\usepackage{multirow}%
\usepackage{amsmath,amssymb,amsfonts}%
\usepackage{amsthm}%
\usepackage{mathrsfs}%
\usepackage[title]{appendix}%
\usepackage{xcolor}%
\usepackage{textcomp}%
\usepackage{manyfoot}%
\usepackage{booktabs}%
\usepackage{algorithm}%
\usepackage{algorithmicx}%
\usepackage{algpseudocode}%
\usepackage{listings}%

\begin{document}

\title[Article Title]{Applications of Superconductor --- Normal Metal Interfaces}


\author*[1]{\fnm{S.A.} \sur{Lemziakov}}\email{sergei.lemziakov@aalto.fi}

\author[1]{\fnm{B.} \sur{Karimi}}\email{bayan.karimi@aalto.fi}

\author[1,2]{\fnm{S.} \sur{Nakamura}}\email{shuji.nakamura@aist.go.jp}

\author[1]{\fnm{D.S.} \sur{Lvov}}\email{dmitrii.lvov@aalto.fi}

\author[1]{\fnm{R.} \sur{Upadhyay}}\email{rishabh.upadhyay@aalto.fi}

\author[1]{\fnm{C.D.} \sur{Satrya}}\email{christoforus.satrya@aalto.fi}

\author[1]{\fnm{Z.-Y.} \sur{Chen}}\email{ze-yan.chen@aalto.fi}

\author[1]{\fnm{D.} \sur{Subero}}\email{diego.suberorengel@aalto.fi}

\author[1]{\fnm{Y.-C.} \sur{Chang}}\email{yu-cheng.chang@aalto.fi}

\author[1]{\fnm{L.B.} \sur{Wang}}\email{libin.wang@aalto.fi}

\author[1]{\fnm{J.P.} \sur{Pekola}}\email{jukka.pekola@aalto.fi}


\affil*[1]{\orgdiv{PICO group}, \orgname{QTF Centre of Excellence, Department of Applied Physics, Aalto University}, \orgaddress{\postcode{P.O. Box 15100}, \street{FI-00076 Aalto}, \country{Finland}}}

\affil[2]{\orgdiv{National Institute of Advanced Industrial Science and Technology (AIST)}, \orgname{National Metrology Institute of Japan (NMIJ)}, \orgaddress{\street{1-1-1 Umezono}, \city{Tsukuba}, \postcode{Ibaraki 305-8563}, \country{Japan}}}

\abstract{The importance and non-trivial properties of superconductor normal metal interfaces was discovered by Alexander Fyodorovich Andreev more than 60 years ago. Only much later these hybrids have found wide interest in applications such as thermometry and refrigeration, electrical metrology, and quantum circuit engineering. Here we discuss the central properties of such interfaces and describe some of the most prominent and recent applications of them.}


\maketitle

\section{Introduction}\label{sec1}

Superconductor-normal metal interfaces form powerful tool for low temperature physics research and applications. They are a basis for a wide range of devices based on clean superconductor/normal metal (SN) contacts and normal metal/insulator/superconductor (NIS) tunnel junctions and more complicated multi-interface structures such as SNIS, SINIS etc. The flow of electrical current and heat through SN interfaces occurs due to single electron tunneling and Andreev reflection {\cite{andreev}} in NIS junctions.

For many years NIS junctions are used for material characterization, for example, measurements of the gap in superconductors. Rapid technology progress in micro-structures fabrication widens the use of SN structures to new areas of research. Practical applications of NIS tunnel junctions are well known in thermometry, electron cooling and low-temperature bolometers. Also, SN interfaces could be extremely useful as frequency and power converters, sensors, and controllable heaters and coolers in quantum electrodynamics and thermodynamics experiments. In this paper we discuss the modern applications of the SN interfaces. 

\section{Thermometry}\label{sec3}

A common application of SN interfaces is local thermometry on a chip. There are several ways to utilize these structures for the purpose of temperature measurement: a tunnel coupled NIS structure (I for insulator), proximity SNIS thermometer, and SNS thermometer. 

\subsection{NIS thermometry}\label{subsec311}

\begin{figure}[ht]
	\centering
	\includegraphics [width=\textwidth] {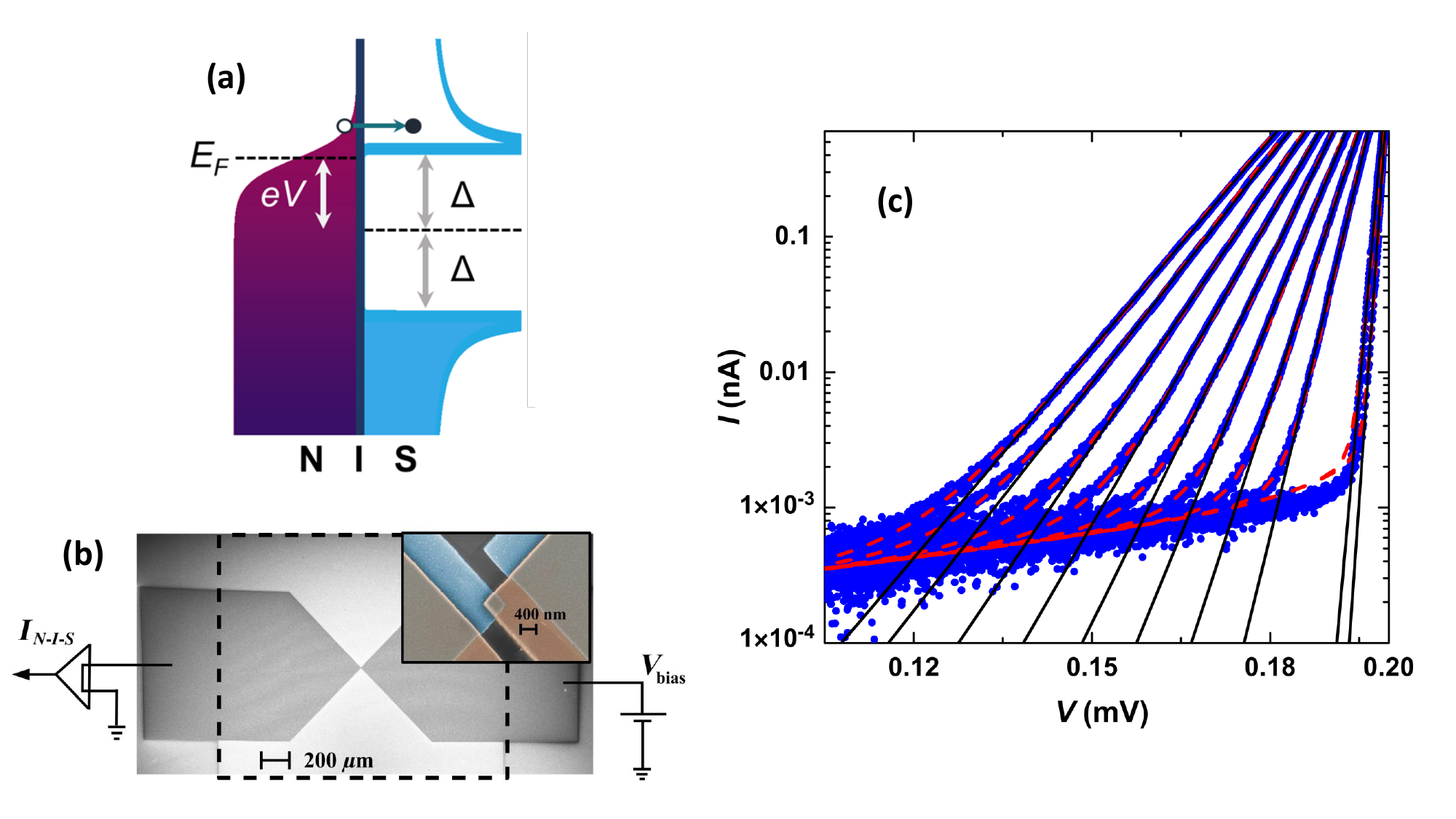}
	\caption{(a) The schematic picture of the DOS and the single electron tunneling at finite temperature in NIS junction with voltage $V$ between the normal metal and superconductor. (b) A scanning-electron micrograph of the NIS-thermometer together with a schematic of the experimental setup. (c) Measured I-V characteristics of the NIS junction (blue dots) at various temperatures and numerical fits as dashed red (from Eq.~\eqref{nisthermo1}) and solid black lines (from Eq.~\eqref{nisthermo2}). The figure is adopted from \cite{Feshchenko2015}.
		\label{NIS}}
\end{figure}

In this sub-section we describe the basic NIS thermometer. Transport through a tunnel junction between a superconductor and a normal metal becomes temperature dependent because the tunneling electrons need to overcome the gap in the density of states (DOS) of the superconductor, Fig.~\ref{NIS}(a). For a sufficiently opaque tunnel barrier, the current $I$ is carried by single electrons, and its voltage $V$ dependence is given by a simple Landauer type expression \cite{Giazotto2006} 
\begin{equation}\label{nisthermo1}
I =\frac{1}{2eR_T}\int n_S(E)[f^N(E-eV)-f^N(E+eV)]dE,
\end{equation}
where $R_T$ is the resistance of the junction, $n_S(E)$ is the normalized DOS of the superconductor, and $f^N(E)$ is the energy distribution of the normal metal. For a conductor at temperature $T_i$, this is given by the Fermi-Dirac distribution, $f^i(E)=(1+\exp (E/k_BT_i))^{-1}$. Equation \eqref{nisthermo1}  then yields the temperature of the normal metal. Note that this expression is sensitive to the temperature of N only, not that of S. Making a S lead in tunnel contact to N thus yields the local electronic temperature of N.

The NIS junction is in principle a primary thermometer in the regime well below the superconducting transition temperature $T_c$. This follows directly from Eq. \eqref{nisthermo1}, which for $T \ll T_c$ and $0 < V < \Delta/e$ ($\Delta$ is the superconducting gap) yields
\begin{equation}\label{nisthermo2}
d\ln(I)/dV = e/k_B T.
\end{equation}
Figure \ref{NIS}(c) shows measurements of the current of a NIS junction on logarithmic scale against voltage at various temperatures demonstrating this dependence \cite{Feshchenko2015}. In practical thermometry this primary property is not commonly used, but the measurement of either conductance or voltage at a fixed bias current serves as a practical thermometer.

Besides the single electron current, Andreev current could be observed in NIS tunneling junctions at low temperatures near zero voltage bias ($V < \Delta/e$) \cite{seliverstov2016, seliverstov2017}. This effect is one of the reasons for the saturation, or loss of sensitivity, for the NIS thermometers at low temperatures. But, at the same time, Andreev current could be utilized as a temperature probe, as demonstrated in Ref.~\cite{faivre2015}. One of the main advantages of Andreev thermometer is its sensitivity also at low temperatures.

\section{Proximity thermometry}\label{subsec313}

Despite numerous commendable features and properties that the basic NIS thermometer has, unfortunately its sensitivity diminishes significantly at lower temperatures. To address this limitation, we have developed a new type of thermometer that operates based on the proximity effect. We add a superconducting lead via a direct metal-to-metal contact (SN) to a NIS junction. This additional superconductor lead induces the proximity effect to the N lead, enabling supercurrent flow through the tunnel junction. This structure is called SNIS, where N now stands for proximitized normal metal. Moreover, the SN contact acts as a heat mirror. By measuring the conductance of the junction as detailed in Ref.~\cite{karimi2020}, it becomes evident that the SNIS proximity thermometer is noninvasive (measured at zero bias) and maintains sensitivity without saturation at low temperatures in a proper setup to well below 20 mK, in contrast to the NIS thermometer. 

\begin{figure}[b!]
	\centering
	\includegraphics [width=\textwidth] {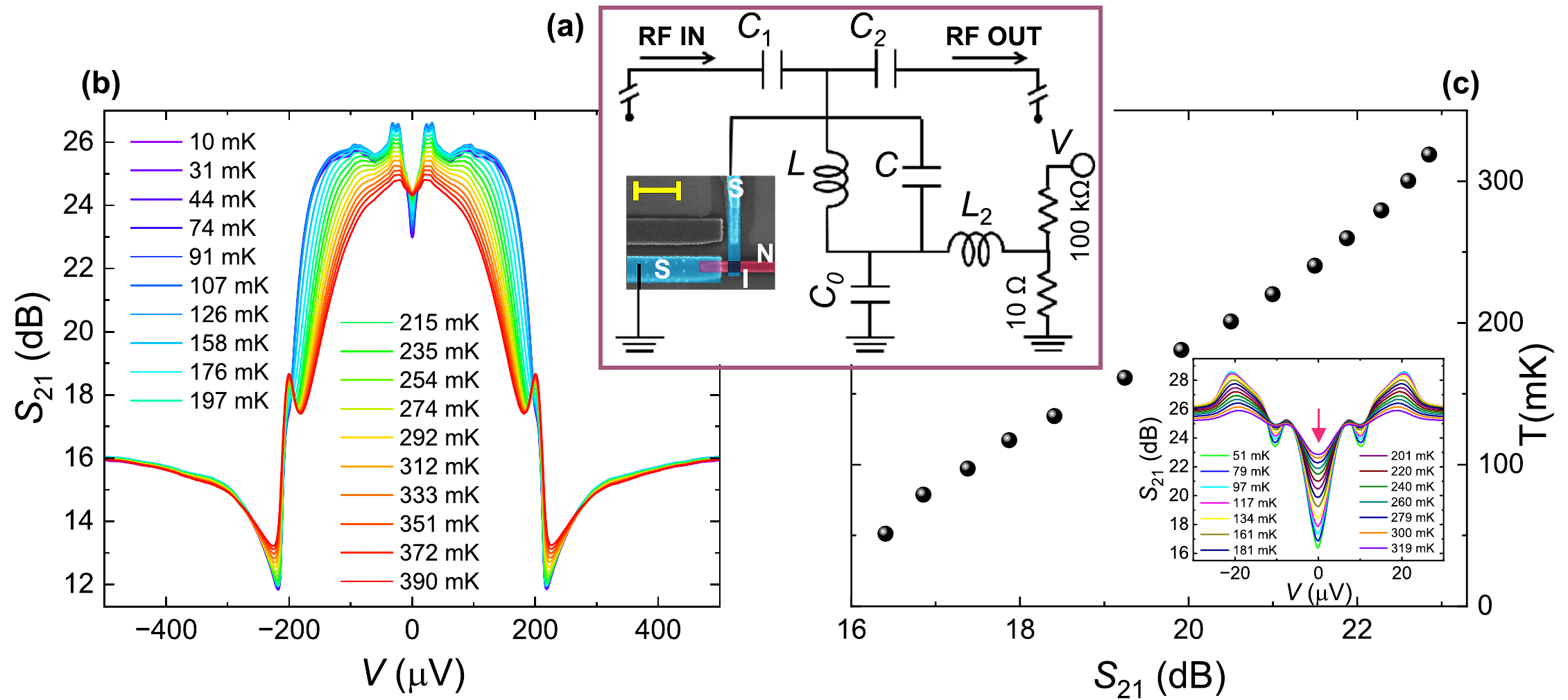}
	\caption{The RF setup for fast temperature measurement with a SNIS junction. (a) The proximity thermometer SNIS structure (S in blue, N in red and I in grey) is embedded in an $LC$ circuit. (b) The measurement of $S_{21}$ (performed at -110 dBm) as a function of bias voltage for a junction at different bath temperatures in the range 10-390 mK. (c) Inset: the same measurement as (b) done at -120 dBm and bath temperature in the range of 50-320 mK. Main panel displays the extracted $S_{21}$ data points of the inset (indicated by the red arrow) at $V=0$ and at different temperatures.
		\label{rf-SNIS}}
\end{figure}

The standard DC setup on Fig.~\ref{rf-SNIS}(b) limits the bandwidth of the measurement to the kHz range due to the high impedance of the junctions (especially in the case of NIS thermometer) and speciﬁcally because of the stray capacitances of the measurement setup. Hence the dc-conﬁguration is not suitable for fast temperature measurement. To overcome this obstacle, we use a RF-setup initially used for NIS thermometry \cite{schmidt2003, gasparinetti2015} with a SNIS probe. For this we embedded a SNIS junction in a distributed $LC$-resonator setup as schematically shown in Fig.~\ref{rf-SNIS}(a). As shown in this panel, the $LC$ resonator is connected to input and output RF lines via coupling capacitors $C_1$ and $C_2$ ($C_1\ll C_2$) and the readout measurement is performed in the transmission mode $S_{21}$ at the resonance of the $LC$ circuit loaded by the SNIS thermometer junction in parallel. The measurement of $S_{21}$ as function of applied dc voltage $V$ at different bath temperatures $T$ is shown in Fig.~\ref{rf-SNIS}(b). The zone of interest for this measurement lies within the zero-bias anomaly range. The inset of Fig.~\ref{rf-SNIS}(c) presents the zoom-out of this range, where we extract the data points of $S_{21}$ at different bath temperatures and a zero-voltage bias (indicated by a red arrow). The main panel of Fig.~\ref{rf-SNIS}(c) shows these extracted points as a function of bath temperature.

Two crucial features of this SNIS thermometer are its sensitivity at low temperatures and its noninvasiveness since it operates at zero bias. Moreover, it has been proven experimentally that this thermometer has reached the ultimate noise level dictated by the fundamental thermal ﬂuctuations \cite{karimi2020det}. These characteristics of this thermal detector enable us to explore the scheme of detecting single microwave photons in a continuous manner, calorimetrically. It is expected to be capable of detecting an absorption event with energy of approximately 1 K × $k_B$, aligning with the range of a photon emitted by, for example, a standard superconducting qubit.

\subsection{SNS thermometry}\label{subsec312}

When a normal metal is placed between two superconductors, a supercurrent can flow through the junction without dissipation due to the Andreev reflection at both SN interfaces. The SNS junction switches to the resistive state once the bias current exceeds the critical switching current $I_c$. Depending on the length of the normal metal $L$, the zero-temperature $I_c$ is determined by either the Thouless energy $E_{th} = {\hbar}D/L^2$, where $D$ is the diffusion constant, or the superconducting gap, $\Delta$, of the superconductor. For the SNS Josephson junction, $I_c$ depends on the electron temperature in the normal metal, thus, it can be used as a thermometer to measure its electron temperature. In the long junction limit with a junction length, $L$, much larger than $\sqrt{{\hbar}D/{\Delta}}$, Dubos et al. experimentally and theoretically investigated the temperature dependence of $I_c$ and found a good agreement between the two \cite{Dubos2001b}.

The dynamics of the JJ can be described using the Resistively and Capacitively Shunted Junction (RCSJ) model \cite{McCumber1968}, where the junction is analogous to a phase particle moving in a washboard potential. Biasing the junction tilts the average slope of the washboard. Once the biasing current exceeds $I_c$, the phase particle runs down the well with a finite voltage drop across the junction. Escaping from the potential well is mediated by macroscopic quantum tunneling and thermal activation. Due to both quantum and thermal fluctuations, the switching of the SNS junction from the zero-resistance state to the dissipative resistive state exhibits stochastic behavior \cite{Fulton1974, Martinis1987, Clarke1988}. 

To measure the distribution of the $I_c$ and determine its value preciously, one can send a train of $N$ current pulses to the junction, at fixed pulse amplitude. The corresponding number of switching events, $n$, is registered by measuring the voltage drop across the junction. The switching probability at a given bias current is $P = n/N$. Subsequently, one can define $I_c$ as the biasing current at which $P$ reaches a defined value. This measurement technique has been previously used to characterize a flux qubit \cite{Chiorescu2003} and investigate break junctions \cite{DellaRocca2007, Zgirski2011}. More recently, researchers have employed the technique to accurately measure $I_c$ of the Dayem nanobridge \cite{Zgirski2018} and SNS junction \cite{Wang2018} for use as a fast thermometer. The time resolution of the thermometer can reach below 10 ns, which is fast enough for it to be used to monitor the thermal relaxation times in the long superconducting wire or in the normal metal at low temperatures. 

The drawback of the SNS Josephson thermometer is its long initialization time after each single-shot measurement and the requirement for many measurements to determine the $I_c$ distribution. For instance, in order to use it as a detector to measure the thermal effect induced by the absorption of a single photon, one must synchronize the photon source and readout detector. The SNS Josephson junction can also be operated as a threshold detector when biased close to $I_c$ in the zero-resistance state. The energy absorbed by the normal metal drives the junction into the resistive state with a voltage drop across it \cite{Revin2020}; in this case, the initialization process does not limit the speed of the single shot measurement, however, it is not possible to determine the amount of energy absorbed by the normal metal.

\section{Cooling using a hybrid NIS junction}\label{subsec32}

In this section, we will explain the mechanism of electron cooling which originated from the elastic single electron tunneling process inherent in NIS junctions. In addition, we will explain another cooling mechanism associated with inelastic electron tunneling with photon absorption, specifically addressing the ``Brownian refrigerator'' \cite{Pekola2007} and the ``Quantum Circuit Refrigerator'' \cite{Tan2017}.

\subsection{Cooling mechanism due to the elastic tunneling}
\begin{figure}[t] 
\centering
\includegraphics[width=13cm]{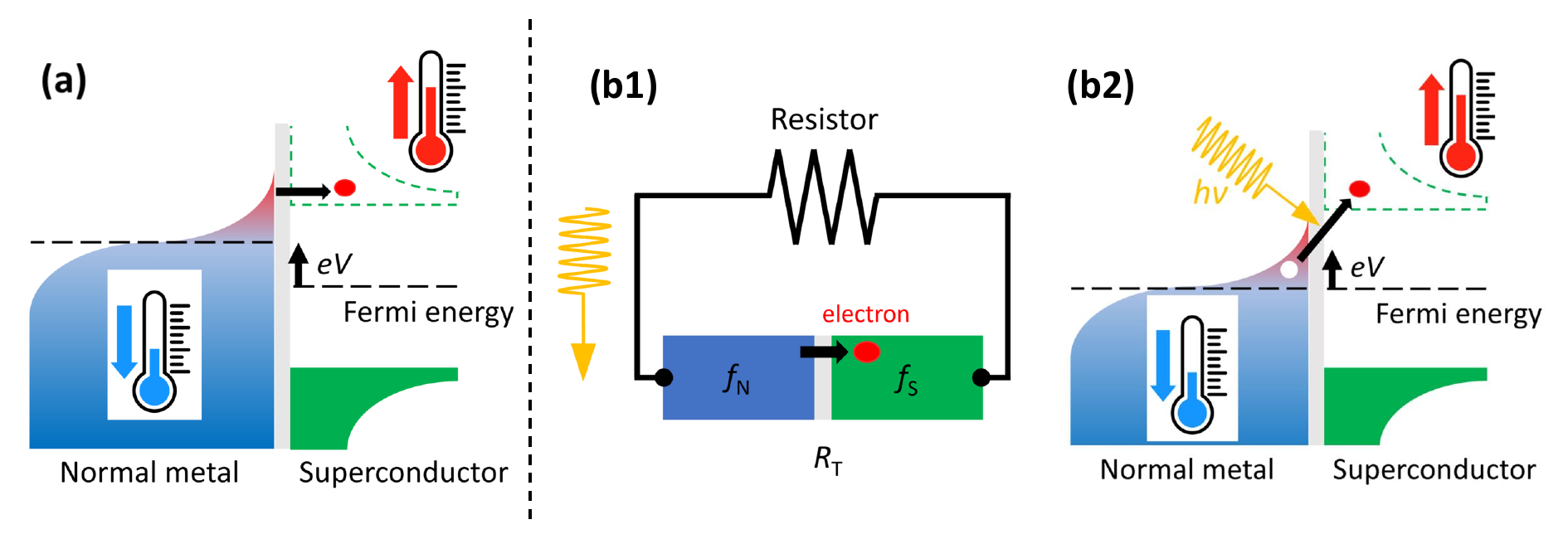}
\caption{\textbf{(a)} Electron cooling at the interface of the hybrid NIS junction. Under the condition of a finite bias voltage $\sim \Delta/e $, hot electrons tunnel into the vacant quasiparticle energy levels of a superconductor, resulting in the cooling of the normal metal. \textbf{(b1)} The schematic picture of the Brownian refrigerator. The thermal noise from the resistor which is connected to the SIN junction is absorbed through the inelastic tunneling process of the electrons inside the normal metal. \textbf{(b2)} In the Brownian refrigerator the temperature of the normal-conducting metal is cooled due to the predominance of higher energy electrons escaping into the superconducting side.}
\label{SIN_cooling}
\end{figure}

The primary mechanism of cooling at the interface of hybrid NIS junction arises from the ``elastic'' tunneling of ``hot'' electrons from the normal metal into the density of states for quasiparticles in the superconductor \cite{Nahum1994}. This type of cooling is called ``electron cooling''. Figure \ref{SIN_cooling} shows the schematic picture of this mechanism. To initiate electron cooling, one must apply a bias voltage $V$ across the NIS junction. Under thermal equilibrium conditions, electron energies within the normal metal are distributed according to the Fermi distribution function. With the application of a bias, electrons of higher energy start to tunnel into the vacant states in the superconductor. Through this tunneling process, lower-energy electrons are left behind in the normal metal, and the electrons subsequently relax to a thermal equilibrium state characterized by a colder Fermi distribution function, due to thermal equilibration processes such as electron-electron and electron-phonon interactions. 

The cooling power of the process (the energy extracted from the normal metal per unit time) can be written by an equation with the similar form as Eq.~(\ref{nisthermo1}) as:

\begin{equation}\label{niscool1}
P_\mathrm{NIS} = \frac{1}{e^2 R_\mathrm{T}}\int dE (E-eV) n_S(E) [f^\mathrm{N}(E-eV) -f^\mathrm{S}(E-eV) ].
\end{equation}

Note that, unlike Eq.~(\ref{nisthermo1}) this expression depends on the temperature of both the superconductor and normal metal. 

Previous studies \cite{Anghel2001} reveal that the optimal bias voltage for maximum electron cooling is $V_\mathrm{optimal} = (\Delta - 0.66k_BT)/e $, and the maximum power is 

\begin{equation}\label{niscool2}
P_\mathrm{optimal} \approx \frac{\Delta^2}{e^2 R_\mathrm{T}}[0.59 (\frac{k_\mathrm{B} T_\mathrm{N}}{\Delta})^{\frac{3}{2}} - \sqrt{\frac{2\pi k_\mathrm{B} T_\mathrm{S}}{\Delta}}\mathrm{exp}(-\frac{k_\mathrm{B}T_\mathrm{S}}{\Delta})],
\end{equation}
where $T_N$, $T_S$ are the temperatures of the normal metal and superconductor, respectively. 

Historically, electron cooling with a hybrid NIS junction was realized by M. Nahum et al. in 1994 \cite{Nahum1994}. Before the cooling in the hybrid NIS junction was reported, it was known that when superconductors with different superconducting gaps (e.g., Nb and Al) were connected via a tunnel junction, the gap of the superconductor with the smaller superconducting gap increased by absorbing a quasiparticle \cite{Blamire1991,Heslinga1993}. M. Nahum et al. converted this superconductor into a normal metal and experimentally demonstrated that electron cooling occurs in the regular N conductor through the aforementioned mechanism. They successfully demonstrated that the electron temperature became lower than the lattice temperature ($\sim$ 100 mK) and the cooling power is roughly 7 fW at 100 mK. After the first observation of the electron cooling, M. Leivo et al. \cite{Leivo1996} demonstrated that more efficient electron cooling with the series connection of two NIS junctions whose resistance is relatively small (around one k$\Omega$) compared to the previous experiment. They demonstrated the cooling power $\sim$ 1.5 pW and reached 100 mK starting from 300 mK. Shortly after these experiments, it was reported that by isolating the normal metal from the larger thermal bath using a membrane, the lattice temperature could also be cool \cite{Manninen1997}. They successfully showed that the temperature of macroscopic size (roughly $0.7 \times 0.7$ mm ) of the thin film membrane (and the normal metal) also can be cooled down. This experiment demonstrated that the hybrid SIN(IS) junction is capable of cooling not only the electronic temperature but also the lattice temperature. This revelation significantly broadened the potential applications of NIS junctions.  

To enhance the cooling power of the hybrid SIN(IS) junction, efforts were concurrently made to identify and address the factors that diminish its cooling power. From Eqs.~(\ref{niscool1}) and (\ref{niscool2}), it is obvious that reducing the resistance of the NIS junction enhances its cooling capability \cite{Leivo1996}. However, such endeavors unveiled certain challenges. Firstly, when trying to reduce the resistance of the tunnel junction by thinning the thickness of the aluminum oxide layer, it becomes challenging to fabricate a high-quality junction. Furthermore, the reduction of the resistance of the tunnel junction leads to the Andreev current $I_\mathrm{AR}$ into the normal metal, and it generates the power $P_\mathrm{AR}= I_\mathrm{AR}\times V$ \cite{Bardas1995}. Furthermore, reducing the tunnel resistance leads to a reduction of the superconducting pair potential at the interface between the normal metal and the superconductor due to the inverse proximity effect. Moreover, with higher frequency and subsequent higher power to S, this back electrode heats up significantly. These phenomena disturb quasiparticle diffusion, and the quasiparticles tunnels back into the normal metal.

\begin{figure} 
\includegraphics[width=13.2cm]{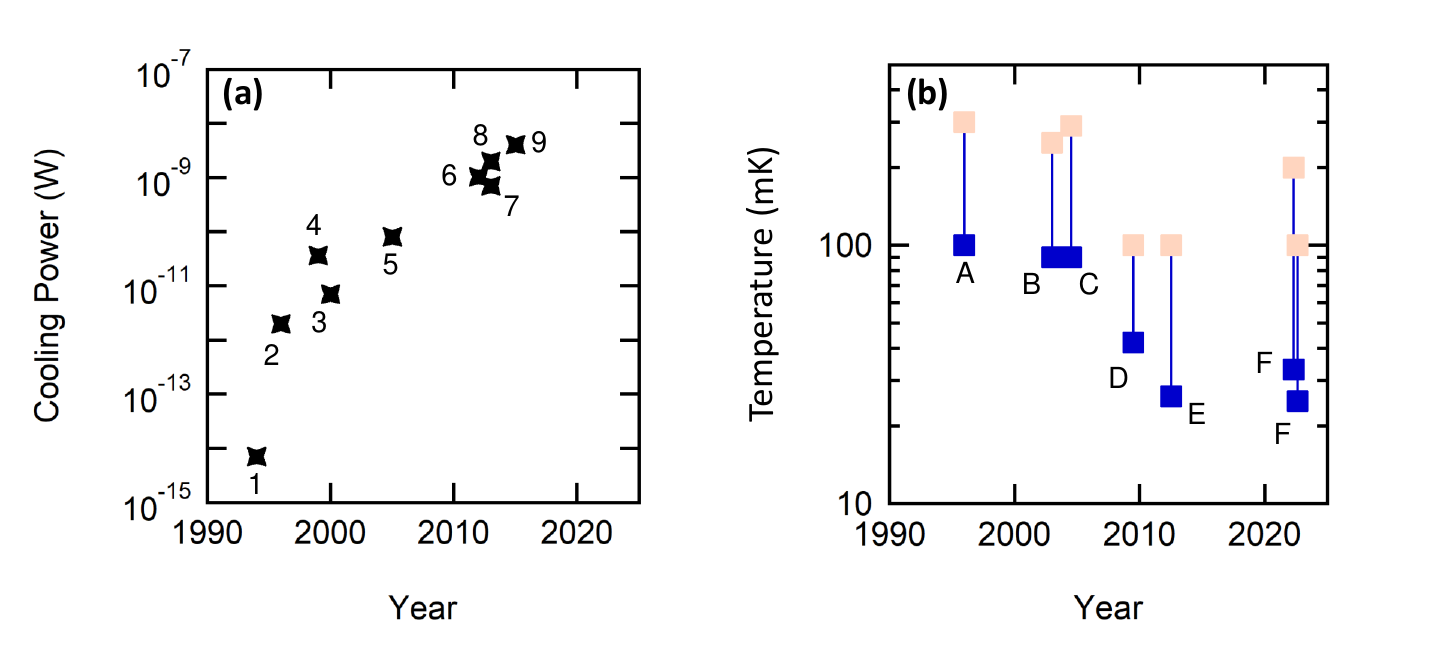}
\caption{Advances in the cooling power \textbf{(a)} and the lowest temperature \textbf{(b)} in the electron cooling with NIS junctions. In the graph \textbf{(b)} faded red squares are the initial temperature at the start of cooling, while blue squares indicate the lowest temperature achieved by the electron cooler. Data in figure \textbf{(a)}: 1 --- Nahum \textit{et al.} \cite{Nahum1994}; 2 --- Leivo \textit{et al.} \cite{Leivo1996}; 3 --- Fisher \textit{et al.} \cite{Fisher1999}; 4 --- Pekola \textit{et al.} \cite{Pekola2000}; 5 --- Clark \textit{et al.} \cite{Clark2005}; 6 --- Nguyen \textit{et al.} \cite{Nguyen2012}; 7 --- Lowell \textit{et al.} \cite{Lowell2013}; 8 --- Nguyen \textit{et al.} \cite{Nguyen_2013}; 9 --- Nguyen \textit{et al.} \cite{Nguyen2015}. Data in figure \textbf{(b)}: A --- Leivo \textit{et al.} \cite{Leivo1996}; B --- Tarasov \textit{et al.} \cite{tarasov2003}; C --- Kuzmin \textit{et al.} \cite{Kuzmin_2004}; D --- Koppinen \textit{et al.} \cite{Koppinen2009}; E --- O'Neil \textit{et al.} \cite{O'Neil2012}; F --- Pimanov \textit{et al.} \cite{Pimanov2022}.}
\label{History}
\end{figure}

To address this issue, a quasiparticle trap was developed \cite{Pekola2000}. This trap, placed close to the hybrid NIS junction, employs a normal metal to absorb the ``hot'' quasiparticles that have tunneled into the superconductor. This mechanism effectively prevents the electrons from undergoing back-tunneling. With the introduction of this quasiparticle trap, it greatly improved to cool down even at bath temperatures below 200 mK. Following the introduction of the quasiparticle trap, efforts were made to cool bulk objects, such as the superconducting sensors \cite{Clark2005, Miller2008}. In 2005, A. M. Clark et al. demonstrated the cooling of 450 \textmu m $\times$ 450 \textmu m freestanding $\mathrm{Si}_{3}\mathrm{N}_{4}$ membrane and 250 \textmu m germanium cube from 300 mK to 230 mK \cite{Clark2005}. Also in 2008, N. A. Miller et al. cooled down a X-ray superconducting transition edge sensor with hybrid junction and reached 185 mK from the starting temperature of 260 mK \cite{Miller2008}.

Effective cooling of electron temperatures to below 100 mK was first reported in 2009 by several groups \cite{Koppinen2009, Muhonen2009}. By utilizing suspended nanowires, they successfully cooled both the lattice and electron temperatures below 100 mK. Afterward, some groups reported electron cooling below 100 mK, starting from more than a few hundred mK \cite{O'Neil2012, Nguyen_2013, Gordeeva2020}.

Figure \ref{History} presents the temporal evolution of (a) the cooling power and (b) the lowest temperature (with the initial temperature before cooling). Over the past 30 years, the cooling capacity has increased by approximately a factor of 1000. Initially, it was challenging to cool down to temperatures of about 100 mK. However, now, electron cooling achieves temperatures as low as 35 mK, starting from 200 mK. Furthermore, the size of the cooled object has also been improved. At first, a few hundred nanometer-sized piece of normal metal was cooled down. However, cooling of bulk thin films such as membranes with dimensions in the hundreds of micrometers have been demonstrated, and currently, efforts are being made to cool objects on the centimeter scale \cite{mykkanen2020, Lowell2013, Zhang2015}.

\subsection{Photon absorption of the hybrid NIS junction}
Beyond the cooling mechanism related to the ``elastic tunneling'' of electrons, hybrid NIS junctions exhibit another key cooling pathway mediated also by the ``inelastic tunneling'' of electrons with photon absorption. In this chapter, we will explain the cooling triggered by the inelastic tunneling process, specifically focusing on ``Brownian refrigeration'' \cite{Pekola2007} and the ``Quantum Circuit Refrigerator (QCR)'' \cite{Tan2017}. Both the Brownian refrigerator and the QCR operate based on a process in which electrons in a normal conductor absorb photons and subsequently tunnel into the energy levels of quasiparticles in a superconductor. Here, we will discuss the theoretical background, history, and application of these two cooling mechanisms.

\subsubsection{Brownian refrigeration}
The concept that inelastic electron tunneling accompanied by photon absorption induces a cooling effect in hybrid NIS junctions was first proposed in 2007 \cite{Pekola2007}. Authors considered a system connecting a resistor with the NIS junction and investigated the process where energy is exchanged through photons. Even in the absence of an applied bias voltage, photons can be absorbed due to the photon-assisted tunneling effect in the normal conductor. Consequently, electrons are capable of tunneling from the normal conductor into the quasiparticle states for the superconductor. By appropriately selecting the temperature, only high-energy electrons in the normal conductor can tunnel, resulting in a cooling effect in the normal-metal conductor. This cooling phenomenon can be written as 

\begin{equation}
\label{Brownian_refrigerator}
P_\mathrm{Brown} = \frac{2}{e^2 R_\mathrm{T}}\int_{-\infty}^{\infty}\int_{-\infty}^{\infty} dEdE' \frac{|E'|}{\sqrt{E'^2-\Delta^2}}E f^\mathrm{N}(E)[1-f^\mathrm{S}(E')] \mathcal P(E-E'),
\end{equation}
where the $\mathcal P(E-E')$ is probability density that the environment (resistor) emits the photon whose energy is $E-E'$. Still, experimental realization of this Brownian refrigeration has not been achieved. 

\subsubsection{Quantum circuit refrigeration}
\begin{figure}[h] 
\includegraphics[width=13cm]{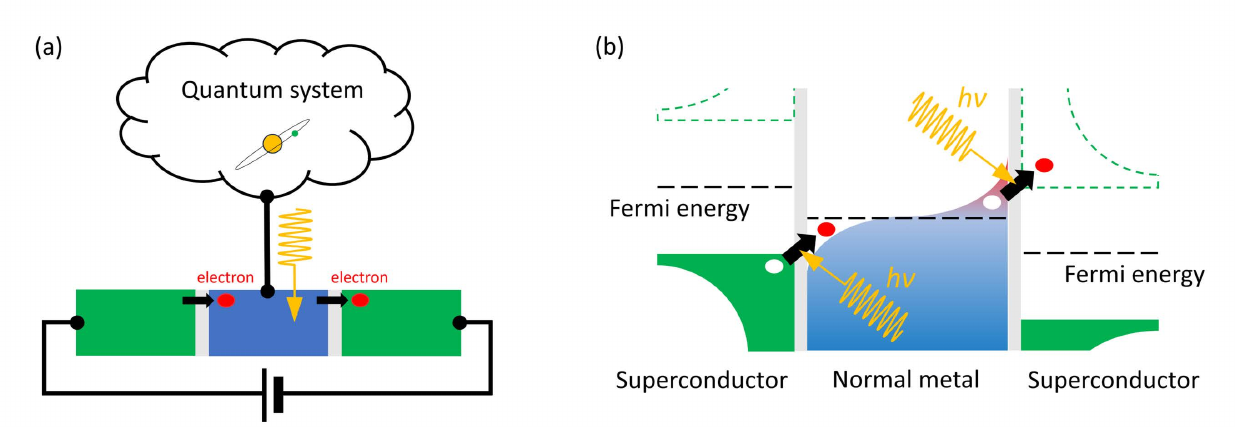}
\caption{The schematic picture of the Quantum Circuit Refrigerator (QCR). (a) The normal metal attached to the system can absorb a photon through the inelastic tunnel process of the electrons. (b) The schematic picture of the process in which the transfer of one electron (quasiparticle) from left lead to right lead with absorption of two photons.}
\label{QCR}
\end{figure}

In 2017, K. Y. Tan and colleagues demonstrated that in a SINIS junction subjected to a bias voltage, the process of inelastic electron tunneling with photon absorption becomes dominant over the process with the photon emission at a certain bias region \cite{Tan2017}. In their first experiment, the normal metal of the SINIS junction was galvanically connected to a coplanar superconducting resonator, leading to a reduction in the resonator's photon number due to the photon absorption of the SINIS. They named this phenomenon ``Quantum Circuit Refrigerator (QCR)''. Figure \ref{QCR} shows the image of the mechanism of this QCR. When a bias voltage around the superconducting gap is applied, photon absorption results in quasiparticle tunneling from the high-bias side of the superconducting lead into the normal metal. Subsequently, electrons within this normal conductor further tunnel into the quasiparticle states in low-bias side of the superconducting lead due to additional photon absorption. So during the process in which the transfer of one electron (quasiparticle) from left lead to right lead, the two photons are absorbed by the QCR. 

The theoretical description for this QCR has been extensively examined by Silveri et al. \cite{Silveri2017}. Following the reported reduction in photon occupation numbers within superconducting resonators achieved through the use of QCR, there have been subsequent reports detailing the achievement of incoherent photon emission and the dynamic control of coupling with external environments using QCR \cite{Mörstedt2022, Sevriuk2022, Yoshioka2023, Masuda2018, Hyyppä2019}. More recently, research has been advancing towards utilizing QCRs for the rapid initialization of superconducting qubits \cite{Sevriuk2022, Yoshioka2023}, with expectations for realizing faster and higher-fidelity initialization. 

\section{Frequency to power and current converters}\label{subsec34}

In this section we discuss the application of NIS-junctions for generating synchronized electric and heat currents. These devices make use of the single-electron charging effect (Coulomb blockade) combined with the gap in the superconducting electrode. The devices are driven by applying a gate voltage at frequency $f$. The charge turnstile produces a current $I=Nef$, where $N$ is an integer \cite{NP2008}. In frequency to power $P$ conversion, the corresponding expression reads $P=2N\Delta f$ \cite{NN2022}.
	\begin{figure*}
		\centering
		\includegraphics [width=\columnwidth] {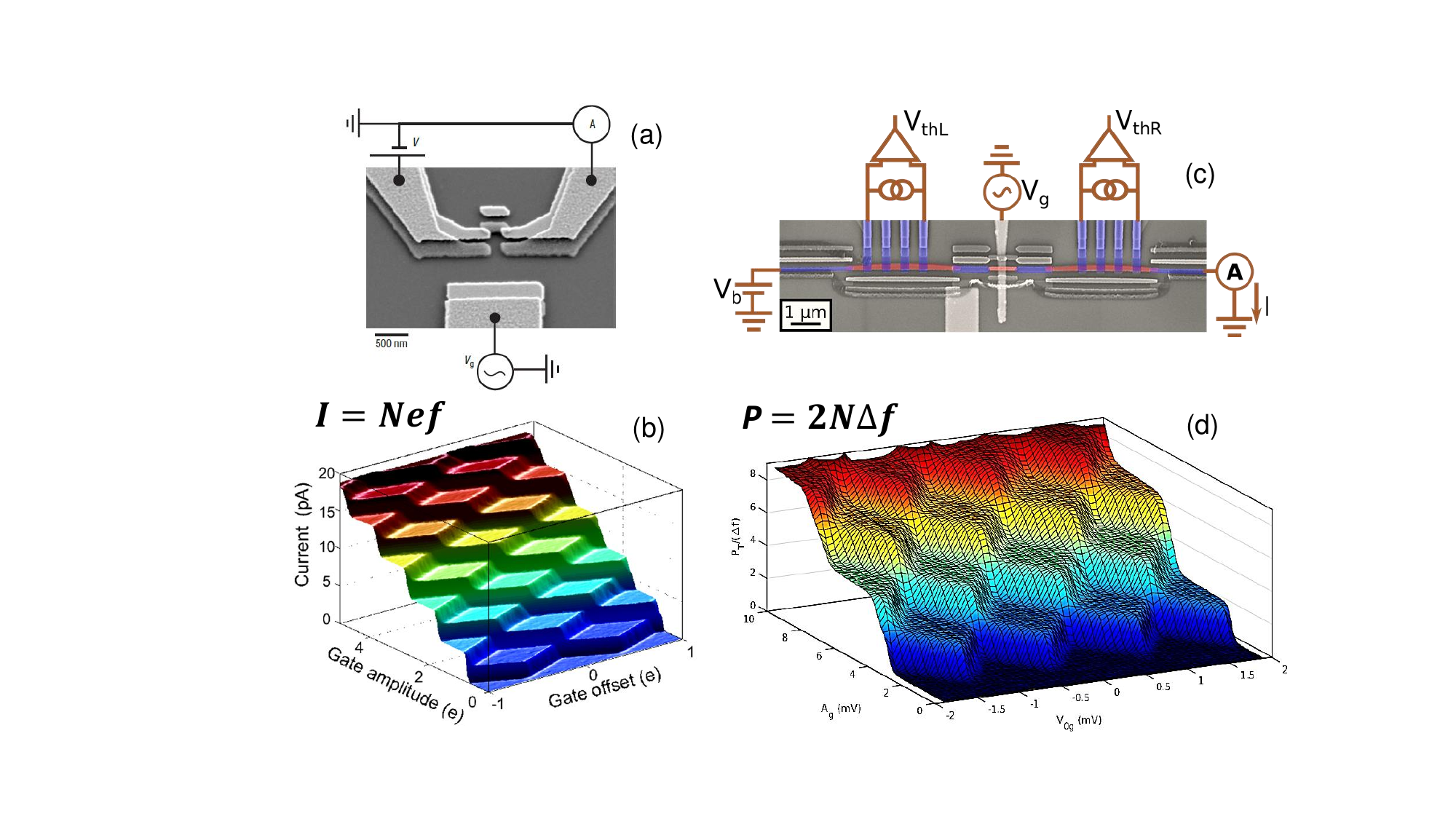}
		\caption{Conversion of frequency to current (ampere) and power (watt). (a) A hybrid single-electron turnstile with normal metal island and superconducting leads, drain-source biased at dc voltage $V$ and driven by a gate voltage with dc bias and ac voltage at frequency $f$. (b) Data of current vs the dc and ac amplitudes of the gate voltage. Drain-source bias is $V \approx \Delta/e$. Plateaus at $I\approx Nef$ are seen. (c) Frequency to power converter. The inner part is as in (a), but the leads include metal absorbers with NIS thermometers on left and right. (d) Data as in (b) but now for the total power absorbed by the left and right ``bolometers''. Here plateaus at $P\approx 2N\Delta f$ of the total power to the two bolometers are seen. The drain-source bias voltage $V=0$ here. Figure adapted from \cite{NP2008} and \cite{NN2022}.	\label{fig_ts}}
	\end{figure*}
	 
\subsection{Ampere: Single-electron turnstile} 
The first realization of an electron turnstile using metallic tunnel junction arrays was put forward more than 30 years ago \cite{geerligs90}. Despite the ingenious idea, those devices never performed accurately enough to be considered seriously for metrological applications. The hybrid single-electron turnstile \cite{NP2008}, on the other hand, is an extremely simple device, consisting of a normal metal island connected to superconducting source and drain electrodes via tunnel junctions, see Fig.~\ref{fig_ts}. When biasing drain-source at a voltage $V\approx \Delta/e$, and by applying an ac gate voltage at frequency $f$ around a fixed dc working point, the device transports a dc current, whose magnitude is very close to the said $Nef$, where $N$ increases stepwise when increasing the ac gate voltage amplitude. 
	
The errors in transferring electrons, deviations in the number transferred per cycle from $N$, arise in the lowest order processes from leakage errors in junctions, and from missing electrons at too fast operation frequency. In principle these errors can be made arbitrarily small by proper design and operation of the device.
	
\begin{figure*}
		\centering
		\includegraphics [width=\columnwidth] {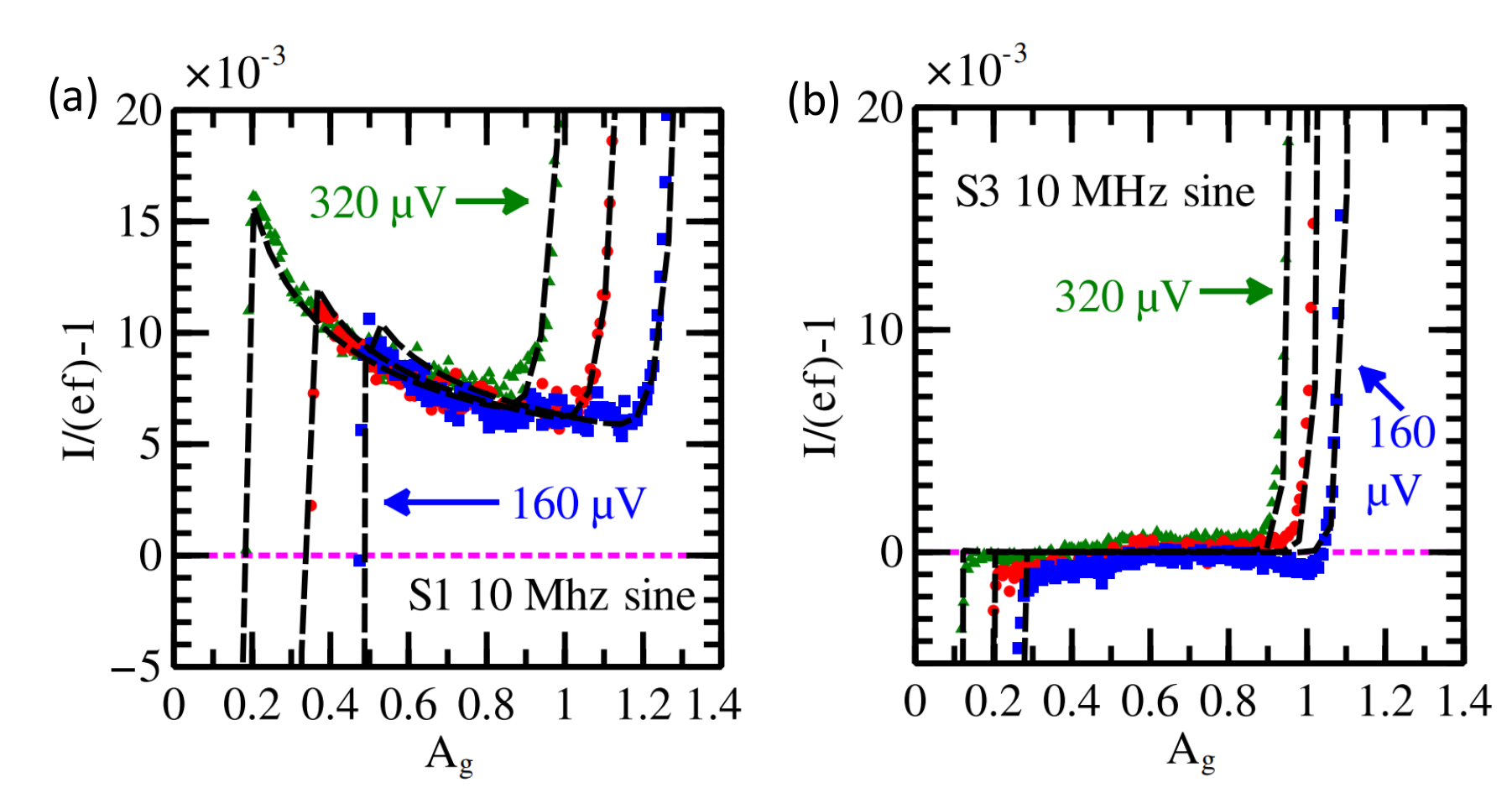}
		\caption{Influence of Andreev tunneling on the transfer accuracy of a hybrid single-electron turnstile. Data of relative excess current with respect to $ef$ on the $N=1$ plateau for two turnstiles, both pumped at $f=10$ MHz, and with various drain-source bias voltages as indicated inside the panels. Sample in (a) has a single electron charging energy $E_C = 0.63\Delta$, whereas the sample in (b) has  $E_C = 1.4\Delta$. The small $E_C< \Delta$ in (a) leads to increased error with the characteristic ac gate amplitude $A_g$ dependence, that can be attributed to Andreev tunneling (dashed lines from the theory), whereas in (b) $E_C > \Delta$ secures accurate pumping. ($A_g$ is normalized by the amplitude corresponding to one gate period.) Figure adapted from \cite{aref2011}.	\label{andreev_epl}}
	\end{figure*}

The more subtle errors in frequency to current conversion arise from Andreev tunneling, where, instead of passing one electron through a junction, the same state is reached by two-electron tunneling accompanied by one electron tunneling in another junction \cite{averin2008}. This leads naturally to an error in the number of transferred charges, i.e. in the current through the device. Influence of Andreev tunneling on current in a turnstile has been quantitatively measured in several works, e.g. in Ref.~\cite{aref2011}, see Fig.~\ref{andreev_epl}. It can be effectively suppressed by increasing the single-electron charging energy of the transistor, i.e. by minimizing the capacitances in the device. The final error in the turnstile current is a third-order tunneling process, which is similar to the Andreev process above, but where the two-electron and one electron tunneling form a quantum mechanical cotunneling process \cite{averin2008}. This error is weak and does not necessarily rule out highly accurate frequency to current conversion, down to relative error rates of $10^{-7}$ or even below, but it cannot be fully eliminated by re-designing the device and parameters.

\subsection{Watt: Frequency to power conversion}
Recently \cite{NN2022} a device with similar architecture was proposed and demonstrated to produce heat currents whose magnitude can be precisely related to the driving frequency. The initial demonstrator is shown in Fig.~\ref{fig_ts}(c). A periodic driving of the gate voltage introduces also heat current in the superconducting leads, since each tunneling event is associated with generation of a quasiparticle with energy $\ge \Delta$. The key difference here as compared to the charge turnstile is that the device can be operated even at zero drain-source voltage, as both outgoing (electron-like) and incoming (hole-like) quasiparticles in the lead produce identical amount of heat. An important ingredient in producing heat current accurately on the level $2N\Delta f$ is that the superconducting BCS density of states has a knife-edge like singularity at $|E|=\Delta$: thus with not too high driving frequency each tunneling electron produces a quasiparticle with this much energy in the lead. The superconducting lead is connected to a normal metal absorber at a distance of few micrometers from the junction. In this configuration, essentially all quasiparticles diffuse to this absorber and release their energy in it, which then converts the quasiparticle energy into heat with practically 100\% efficiency. This method provides currently a \% level accurate source of power in the aW - fW regimes, see Fig.~\ref{fig_ts}(d). The main source of error is of thermal origin, yielding on average an equipartition-like $k_BT/2$ excess power on top of $\Delta$ even at low driving frequencies \cite{prl2022}.  

\section{Quasipartical traps}\label{subsec35}

In this section, we overview the quasiparticle trap as one of the applications of SN interfaces. Superconductors, having zero electrical resistance, are expected to ideally have no energy dissipation. However, in actual devices, the energy such as the thermal disturbances can break Cooper pairs, leading to quasiparticles that cause dissipation. Even more importantly for devices in this article, one injects quasiparticles to the superconductor when biasing the junctions. These quasiparticles are a source of decoherence in quantum bits, reduce the cooling effect in NIS junctions, and lower the quality factor in superconducting resonators. This phenomenon, known as ``quasiparticle poisoning'', is one of the main obstacles to realize superconducting quantum devices \cite{Pekola2000, Karzig2021, Aumentado2023, Patel2017}.

One approach to resolving the issue of quasiparticle poisoning involves the use of ``quasiparticle traps'', which absorb the generated quasiparticles from the superconductor \cite{Pekola2000}. This trapping can be implemented by attaching a normal conductor to a superconductor. When a normal conductor is connected to a superconductor, the proximity effect occurs, where the superconducting order parameter permeates into the normal conductor. Simultaneously, the order parameter of the superconducting side near the junction has a spatial variation, decreasing towards the normal metal. This spatial gradient of the order parameter acts as a potential for quasiparticles, causing them to diffuse towards the normal conductor. In the normal conducting metal (such as copper), the high thermal conductivity at very low temperatures, due to the presence of numerous electrons at the Fermi surface, enables rapid diffusion of the quasiparticle energy into the normal conductor. This quasiparticle trap has been utilized in various superconducting devices, such as NIS coolers \cite{Pekola2000,Peltonen2011}, SINIS turnstiles \cite{Taupin2016, Nakamura2017}, superconducting qubits \cite{Wang2014} and in semiconductor and normal metal hybrid devices \cite{Sato2022}.

\begin{figure}[ht] 
\centering
\includegraphics[width=6cm]{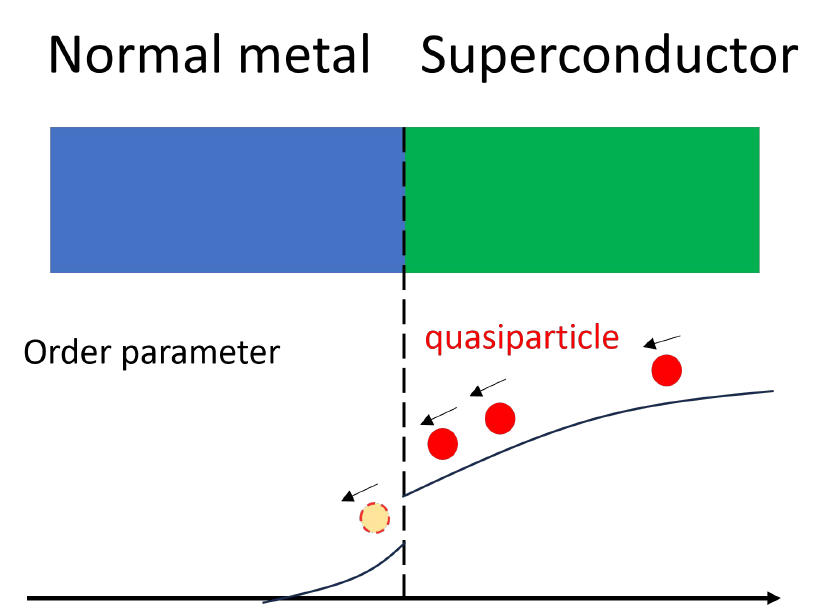}
\caption{Quasiparticle trap at the interface between the superconductor and the normal metal. The pair potential of the superconductor at the interface between the normal metal and the superconductor is suppressed by the inverse proximity effect. The spatial gradient of the pair potential accumulates the quasiparticles towards the interface.}
\label{Quasiparticle_trap}
\end{figure}

\section{SN interfaces in quantum thermodynamic experiments}\label{subsec36}

The potential of devices leveraging superconductor-normal (SN) interfaces becomes evident through their implementation in mesoscopic devices designed for heat management. They have proved to be exceptionally valuable for addressing heat dissipation challenges in circuits of the quantum era. Well engineered devices capitalize the unique properties of SN interfaces, with precise control over thermal flow at microscopic scales. Furthermore, the exploration of heat transport in ultra-strong reservoir-resonator and resonator-qubit coupling regimes under the framework of quantum thermodynamics is promising \cite{ Sai, Kosloff, B.Karimi}, however experimental observations remain fewer.

Research in the field of superconducting quantum circuits has evolved significantly in past few decades, promising enormous potential towards their realization in plethora of applications. Particularly in the circuit quantum electrodynamics (c-QED), the architecture incorporating superconducting quantum bits coupled to one or many superconducting resonators, is of paramount interest in realizing diverse novel quantum devises \cite{G.Wendin, XiuGu, JohnClarke, T.P.Orlando,J.E.Mooij,M.H.Devoret}. The fundamentals towards the realization rests in the reliable, robust and most importantly controllable interaction between the quantum bits and superconducting cavities. Various methods to couple superconducting quantum bits to the electromagnetic environment have been proposed and demonstrated in the last decade \cite{Abdumalikov, Wallraff.A, Devoret.M}.

\subsection{Photonic heat transport}\label{subsec361}

Implementation of the tunneling NIS junction into superconducting microwave circuits makes it possible to study experimentally heat transfer mechanisms in the quantum limit, i.e. circuit quantum thermodynamic (QTD). 

At low temperatures, when the electron-phonon and normal electronic heat conduction is highly suppressed, the heat in the superconducting microstructures is transferred by photon radiation. In Ref.~\cite{Meschke2006} it was shown that the thermal conductance between two small metal islands mediated by photons approaches the single-mode quantum limit at low temperatures. Also, the quantum limited heat conduction by photons was experimentally observed over macroscopic distances \cite{partanen2016}. The crossover between photonic and quasiparticle heat fluxes in a superconductor was observed in Ref.~\cite{Timofeev2009}, where the quantum-limited electronic refrigeration was studied.

    \begin{figure*}
		\centering
		\includegraphics[width=\columnwidth]{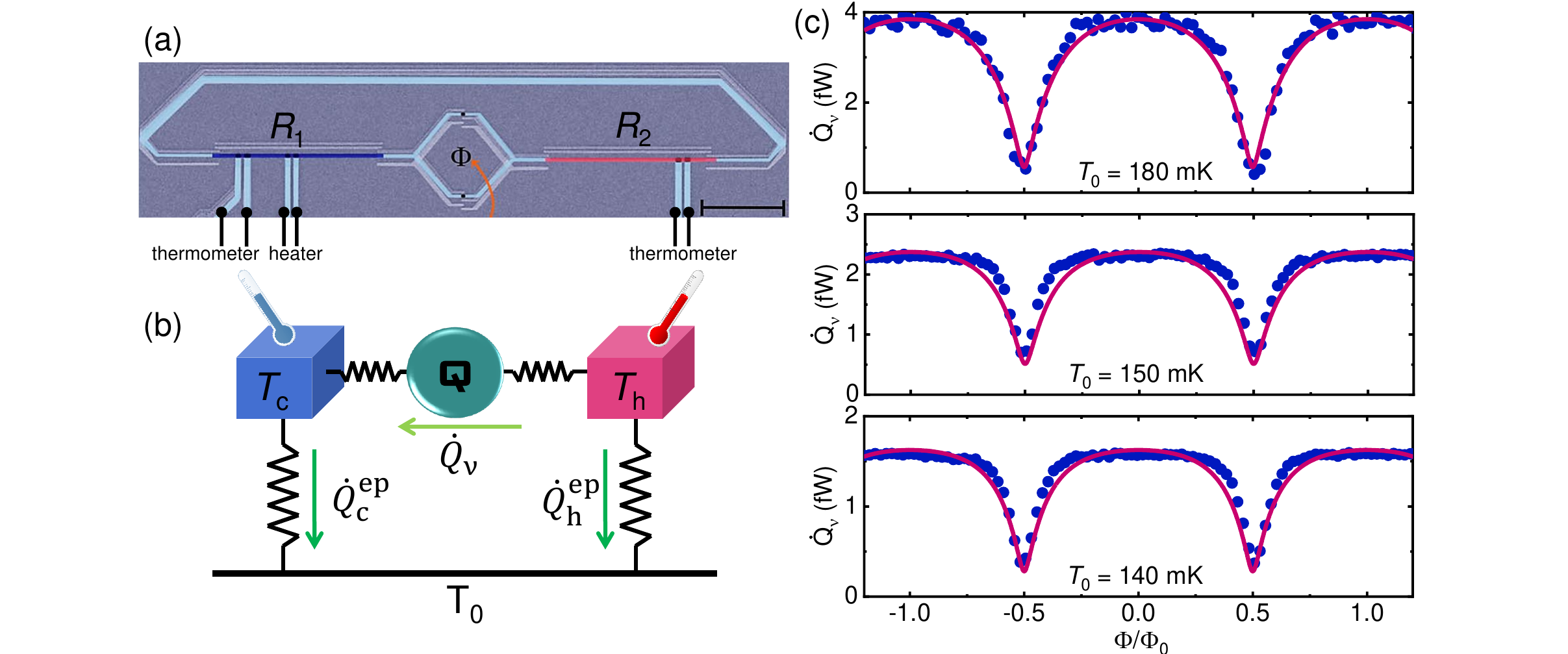}
		\caption{Tunable quantum heat transport. (a) Scanning electron micrograph of a device (scale bar: $5\,\mu{\rm m}$) where two nominally identical normal-metal resistors $R_1$ and $R_2$ made of AuPd (colored in blue and red on the left and right, respectively), are connected to each other via aluminum superconducting lead (light blue) and interrupted by a dc SQUID. Here the SQUID serves as a thermal switch between the resistors and can be controlled by external magnetic flux $\Phi$. The NIS probe junctions (vertical aluminum leads connected through an oxide barrier to the AuPd normal metal) are responsible to heat/cool and monitor the electronic temperature of the normal metals. (b) Schematic illustration of the thermal model of the device shown in (a). The SQUID in the middle regulates the heat flow $\dot{Q}_\nu$ between the two resistors where each of them is coupled to the phonon bath at fixed temperature $T_0$. (c) The SINIS pair in the left resistor is adjusted at its maximum cooling bias and then the heat flows from the right reservoir to the left one. This panel shows the measured heat current by photons as a function of normalized applied magnetic flux $\Phi/\Phi_0$ for three $\dot{Q}_\nu$ different values of the bath temperature $T_0$, where $\Phi_0=h/2e$ is the superconducting magnetic ﬂux quantum. The solid lines are the theoretical results obtained within the linear model by assuming a lumped element approximation, which is valid since the radiation wavelength at temperatures of interest for heat transport measurement is much greater than the circuit dimensions. The experimental parameters: the critical current of the SQUID, $I_C\sim 18$\,nA, $R_1=R_2\sim 1\,{\rm k}\Omega$, the electron-phonon coupling constant of the AuPd, $\Sigma_{\rm AuPd}=3\times 10^9  {\rm WK^{-5}m^{-3}}$.
        \label{Diego_AuPd}}
	\end{figure*}

In Ref.~\cite{partanen2018} a flux-tunable heat sink was implemented. In this experiment the photons from high quality resonator were effectively dissipated via tuning the frequency of the coupled low-quality factor resonator. In Ref.~\cite{Ronzani2018} the tunable photonic heat transport between two normal-resistors was studied. Here resistors with NIS junctions play a role of the controllable thermal reservoirs. Each resistor was embedded into separate coplanar waveguide resonators. The two resonators were coupled to one superconducting transmon qubit, which allowed to tune the heat flow from one resistor to another by flux-tuning of the qubit frequency. It was shown and described theoretically, that in this device (quantum heat valve) the photonic heat transport highly depends on the interplay between the dissipation in the reservoir and qubit coupling.

In the experiments described above, normal metal lead in NIS junctions was made of copper, consequently, thermal bath impedance was relatively small, in order of ohms. By using others materials, like gold-palladium (see Fig.~\ref{Diego_AuPd}) or chromium, it is possible to increase the normal metal resistance up to kOhms. In Ref.~\cite{subero2023} chromium resistors were utilized for experiments on heat transport where a Josephson junction acts as an inductor in the presence of this highly resistive environment. In this work a SQUID was embedded in Johnson-Nyquist setup  with two chromium resistors with nominally equal resistances around 11 kOhm, which is higher than the resistance quantum $R_Q = h/4e^2 =$ 6.45 kOhm. For the DC charge transport measurements the critical current completely vanishes in this regime in agreement with theory \cite{schmid1983, bulgadaev1984, ingold1992}. However, the thermal transport measurements demonstrate clear oscillations of the heat current through the SQUID with magnetic flux, which shows the inductive response of the Josephson junction even in the presence of high impedance environment.

\subsection{Thermal diode}\label{subsec362}

The practical implementation of SN interfaces realized as mesoscopic structures in quantum heat devices finds another interesting application by breaking the symmetry of the structure. The on-chip asymmetry is harnessed by coupling the photonic heat channels to the SN structures with different coupling energies, demonstrating remarkable heat current rectification through the device as a function of the applied flux via a superconducting qubit by measuring the flow of heat current in forward and reverse directions.

Non-linearity, an intrinsic property of superconducting qubits, plays a role that is not only intriguing to fundamental research but also holds promise for practical applications in cryogenic environments \cite{Clarke2008, M.H.Devoret}. The potential of this dynamic property may lead to practical applications, particularly in harnessing devices based on non-reciprocal behavior. In cryogenic read out channels, commercial microwave components are a crucial part, employed to protect the device under test from amplifier backaction and black-body radiation. Commercially available non-reciprocal components leave considerable footprint in dilution cryostats, which significantly restricts the scalability of quantum circuits \cite{A.R.Hamann, ToshiroKodera, GiovanniViola}. Moreover, these available components are based on ferrites, detrimental for quantum circuits \cite{A.R.Hamann, LeonardoRanzani, ArchanaKamal, Zhang}. Recently, researchers have found several alternatives that are both on-chip and ferrite free, progressing towards the commercialization in the near future \cite{Yaakobi, Lecocq, Brien, ShoMasui, Mahoney, Stace, Mason, S.Barzanjeh, Rohit}. Within quantum thermodynamics, heat rectification by an artificial atom is a hot topic useful for management of heat dissipated from superconducting quantum circuits. Recently, the idea of heat rectification by an artificial atom has progressed from theoretical studies to interesting experimental investigations \cite{Giazotto.F, Ordonez, Barzanjeh.S, Kosloff.R, Robnagel, Ronzani2018}. Using a transmon-type two-level system coupled asymmetrically to two resonators with identical frequencies, the implementation of heat rectification was tested \cite{Jorden2020}. The system exhibits both non-linearity and structure asymmetry, crucial to implement non-reciprocity in a device by breaking symmetry. To characterise the associated heat transport, in Ref.~\cite{Jorden2020} , the authors used standard NIS thermometers and heaters as in \cite{Ronzani2018}. The reported heat rectification is a function of applied magnetic flux, and is tunable between 0 and $10 \%$. Another recently published work reports transmission diode effect, where transmission rectification ratio $R$ exceeds $90\%$ when the qubit and resonator are tuned to resonate with each other with the application of magnetic flux \cite{Upadhyay}. In future this flux-qubit based device could be tested in thermal environment using the same NIS-based temperature measurement and control.

\subsection{Controllable thermal bath for qubit thermometry}\label{subsec363}

In recent years the idea of utilizing a qubit as a thermometer has attracted a lot of attention because of both high sensitivity of qubits to noise (see, for example, \cite{Schoelkopf2003, Krantz2019}) and prospects of exploitation of quantum thermometry algorithms \cite{Mehboudi_2019, Jevtic2015, Mukherjee2019}. Experimental realizations of thermometry using QED circuits based on superconducting qubits can be divided into systems sensitive to photons emitted by a qubit \cite{Scigliuzzo2020} or various measurements of qubit population distribution \cite{Geerlings2013, Jin2015, Kulikov2020, Sultanov2021}.

\begin{figure}[b!] 
\centering
\includegraphics[width=8cm]{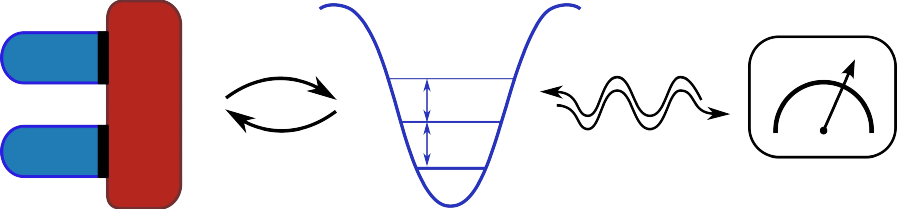}
\caption{Concept of the possible qubit thermometry realization. From the left to the right: controllable heater (SINIS-structure), filtering resonator, and qubit as a sensor.}
\label{QT_concept}
\end{figure}

A qubit as a thermometer could be sensitive to substrate phonons, quasiparticles present in the superconductor \cite{Martinis2009, Catelani2011, Vool2014, Gustavsson2016, Serniak2018}, external radiation field \cite{Goetz2017, Liu2024} and other noise types, such as magnetic flux noise, mechanical vibrations \cite{kono2023}, etc., many of which are temperature dependent. Therefore it is crucial to calibrate the qubit thermometer's response to controllable local on-chip heating, rather than to the whole cryostat heating. A NIS tunnel structure on a normal reservoir is a natural candidate for the role of a controllable environment.

A concept for possible realization of a qubit-based thermometer, sensitive to temperature of another on-chip mesoscopic structure is shown in Fig.\,\ref{QT_concept}. The circuit is comprised of a superconducting transmon qubit which is weakly coupled to a copper mesoscopic normal-metal resistor having controllable temperature via a coplanar waveguide (CPWG) resonator, far detuned from the qubit frequency. NIS-contacts to the resistor are made to adjust and measure the effective electron temperature. Dispersive qubit readout is done using another CPWG $\lambda/4$-resonator.

A few of the transmon parameters are sensitive to the effective temperature of the qubit $T_q$. Firstly, its population distribution $p_i(T_q)$ follows Maxwell-Boltzmann distribution 

\begin{equation}
    p_i=\frac{\exp{\left(-E_i/kT_q\right)}}{\sum_i \exp{\left(-E_i/kT_q\right)}}
    \label{eq.Maxwell-Boltzmann}
\end{equation}

\noindent with qubit level energies $E_i$, $i \in \{g,e,f, ...\}$ . Also the dephasing rate $\gamma_\varphi$ depends on the average photon number $n$ in the resonator  \cite{Schoelkopf2003,Goetz2017}.

\begin{figure}[b!] 
\centering
\includegraphics[width=8cm]{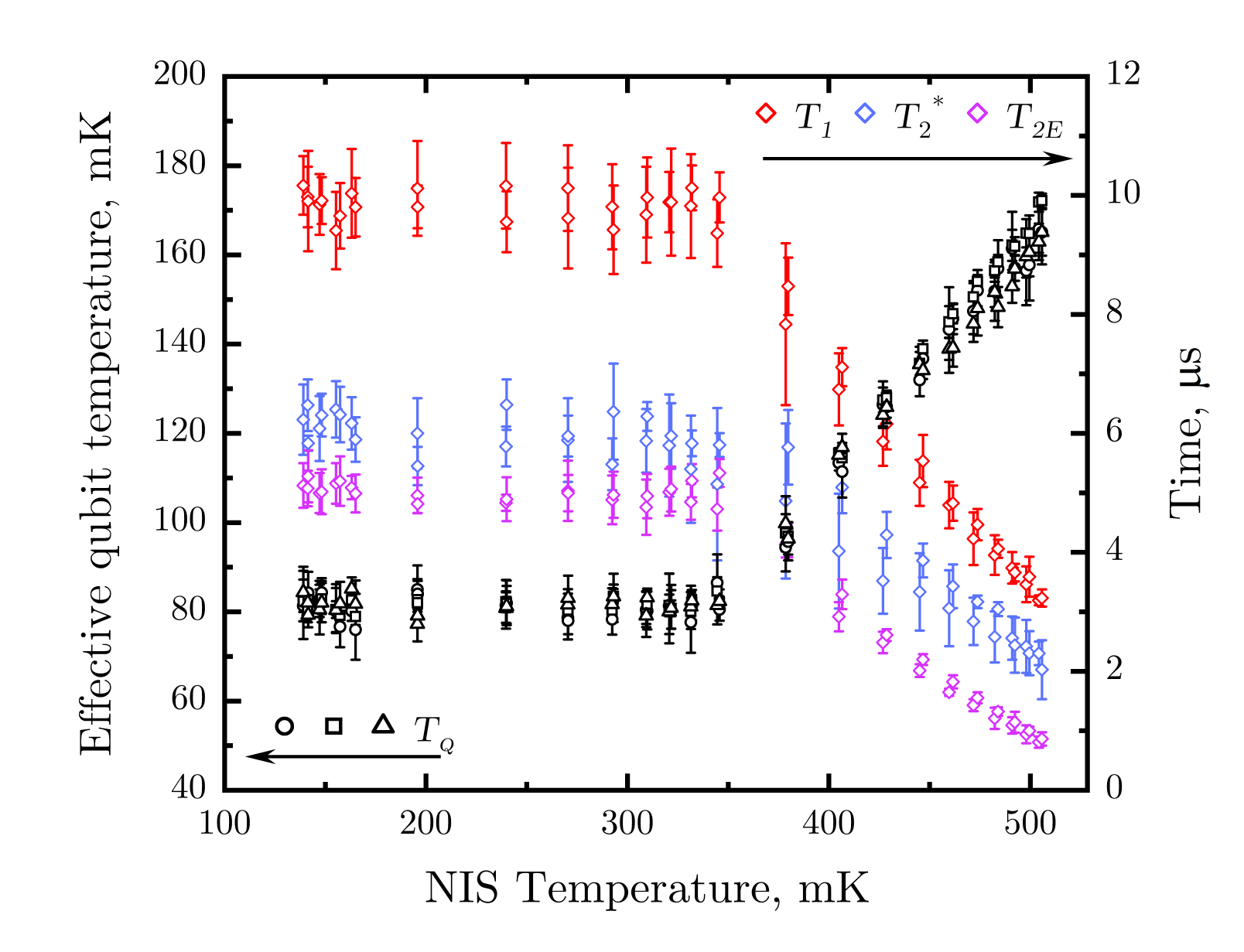}
\caption{Effective qubit temperature $T_Q$ (black) and qubit lifetimes $T_1$, $T_2^\ast$, $T_{2E}$ (red, blue and purple, correspondingly) as functions of the resistor temperature, measured with NIS-thermometer at the base cryostat temperature of 19\,mK. Symbols for $T_Q$ denote different ways of qubit temperature measurement (see \cite{Sultanov2021}).}
\label{QT_data}
\end{figure}

Following this idea, we measured qubit energy relaxation time $T_1$, Ramsey decay time $T_2^\ast$, echo decay time $T_{2E}$ and population distribution $p_i, i\in\{g,e,f\}$ for different heating powers at the resistor, simultaneously measuring its temperature $T_R$ (see Fig.\,\ref{QT_data}).
Population distribution measurements for qubit states $\left|g\right>, \left|e\right>, \left|f\right>$ were done by applying various sequences of $\pi_{ge}$ and $\pi_{ef}$ pulses, with subsequent state readout. The method is described in detail in \cite{Sultanov2021}.

In this experiment below $T_R=320$\,mK the qubit lifetimes are stable and practically insensitive to the power dissipated by the resistor, the qubit population behaves similarly and corresponds to the qubit effective temperature $T_Q\approx80-85$\,mK. At higher resistor temperatures the  qubit lifetimes linearly drop with temperature, while $T_q$ grows, reaching 160\,mK at 500\,mK at $T_R=500$mK. 

In this setup the qubit and the resistor are connected by two energy channels --- photon channel through the CPWG resonator and phonon channel through the substrate. Spectral density of the Johnson-Nyquist noise $S_V(\omega, T_R)$ generated by the resistor changes linearly with $T_R$ in the 100--500\,mK range, which leads to the same linear behaviour of the photon population in the resonator, whereas the electron-phonon coupling follows the power law $P_{e-ph}=\nu\Sigma \left(T_{e}^5 - T_{ph}^5\right)$, where $\nu$ is the volume of normal metal and $\Sigma$ is the electron-phonon coupling constant. Thus, $P_{e-ph}$ quickly grows at higher electron temperatures $T_e$ of the resistor with the phonon temperature $T_{ph}$ stabilized close to the cryostat base temperature. Hence, the heat flowing across the substrate through the predominant phonon channel causes higher quasiparticle density in the vicinity of the qubit, presumably leading to enhanced relaxation and dephasing \cite{Martinis2009, Catelani2011, Vool2014, Gustavsson2016, Serniak2018}.

The measurement data presented in Fig. \ref{QT_data} show that a NIS structure at higher powers of heating cannot be considered as a local heater in a QED circuit. However, sensitivity of the qubit to the local photonic contribution can be increased with stronger coupling between the qubit and the resistor, which remains to be proven experimentally.

\subsection{Heat engines and refrigerators}\label{subsec364}

Experimental realization of quantum heat engines and refrigerators is important for advancement of the field of quantum thermodynamics and quantum technology. It leads us to understand the fundamental relationship between quantum mechanics and thermodynamics \cite{annurev-physchem-040513-103724,Alicki_1979,PhysRevE.76.031105,Binder2019,vischi2019, alicki2023}, and it might significantly contribute to the progress of developing of quantum technology. A fundamental question such as whether quantum coherence can boost the engine performance or not might be experimentally addressed. A multitude of experimental platforms have been proposed and realized to implement quantum thermal machines \cite{Rossnagel325, Josefsson2018, PhysRevLett.123.240601, PhysRevLett.125.166802, germanese2022}. In addition, extensive theoretical operation of quantum thermal machines has been proposed for refrigerators \cite{Abah_2016, PhysRevE.93.062134,PhysRevB.76.174523, PhysRevB.94.184503,PhysRevB.100.085405,PhysRevB.100.035407} and heat engines \cite{PhysRevB.93.041418, Campisi2016, PhysRevE.76.031105, PhysRevB.101.054513, scharf2020}.

\begin{figure*}[b!]
        \center
	    \includegraphics[width=\textwidth,height=\textheight,keepaspectratio]{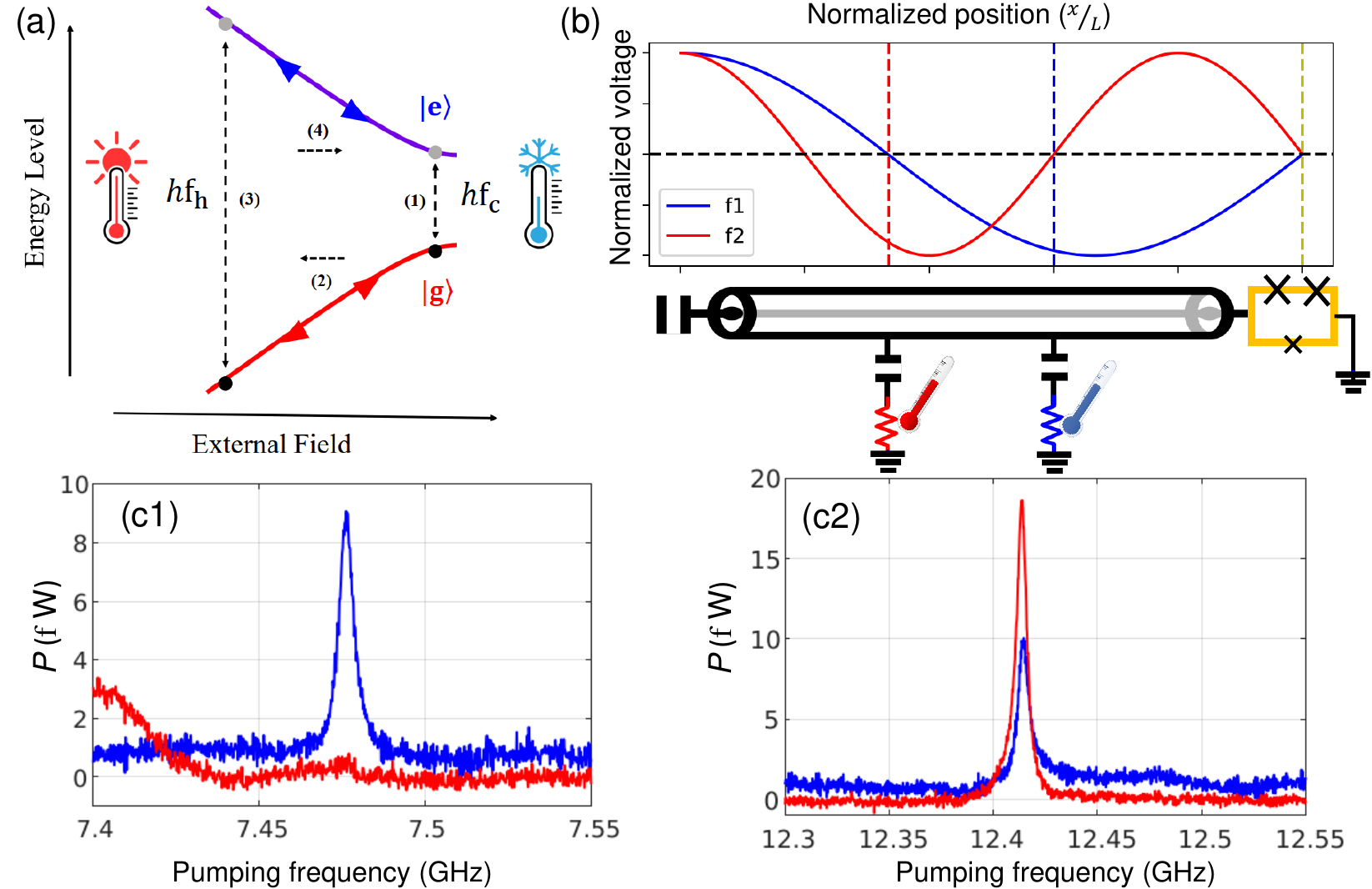}
	    \caption{\textbf(a) Quantum Otto cycle consists of a cyclical interaction of a two-level system with a cold ($f_c$) and hot ($f_h$) bath. It has four strokes: (1) qubit thermalization with the cold bath at frequency $f_c$, (2) an adiabatic stroke of the qubit frequency from $f_c$ to $f_h$, (3) thermalization with the hot bath at frequency $f_h$, (4) an adiabatic stroke from frequency $f_h$ to $f_c$, and back again to (1) for thermalization with the cold bath at frequency $f_c$. External work $W$ is supplied to modulate the qubit energy level. At certain conditions, the heat is pumped from cold to hot bath. \textbf(b-c) A proposal and testing of an experimental realization of an Otto refrigerator in a single superconducting cavity coupled to a qubit. A flux qubit with three junctions coupled at the shorted end of a $\lambda/4$ resonator \cite{R.Upadhyay}. The first and second resonance modes have a voltage distribution along the cavity. \textbf{(b)} The two modes $f_1$ and $f_2$ are coupled independently to a resistor via a capacitor. Here, the coupled $f_1$  mode-resistor is the cold bath while the coupled $f_2$ mode-resistor is the hot bath. The $f_0$ fundamental mode contributes to the background between the two resistors and is set to be far detuned from the qubit frequency. Powers dissipated in the two resistors due to cavity pumping are shown around $f_1 \sim 7.47~\mathrm{GHz}$ \textbf{(c1)} and $f_2 \sim 12.42~\mathrm{GHz}$ \textbf{(c2)} resonances. Blue lines indicate the power to the 'blue' resistor while red lines indicate the power to the 'red' resistor. 
	    \label{fig:1}}
\end{figure*}   

An Otto cycle refrigeration in a superconducting circuit platform is extensively explored and practically achievable for the implementation of a quantum refrigerator \cite{PhysRevB.94.184503,PhysRevB.76.174523,PhysRevB.100.085405,PhysRevB.100.035407}. A quantum Otto cycle consists of sequential interactions between a qubit and a cold and hot reservoir at frequencies $f_c$ and $f_h$, respectively, and it has four strokes as illustrated in Fig.~\ref{fig:1} (a). First, at frequency $f_c$, qubit is in resonance with cold bath and its population is determined by the temperature of the cold bath due to the thermalization. Second, an adiabatic stroke of the qubit frequency from $f_c$ to $f_h$ is performed. Third, thermalization with the hot bath at frequency $f_h$, determines the qubit population by the temperature of the hot bath. Fourth, an adiabatic stroke from $f_h$ to $f_c$ is performed; and finally the qubit is back to the initial state at thermalization with the cold bath at frequency $f_c$. Work $W$ by an external field is applied to cyclically modulate the qubit frequency between $f_c$ to $f_h$ and eventually at steady state to pump heat from the cold to the hot bath. Cooling the cold bath is achieved under the condition $f_h/f_c > T_h/T_c$. 

This cycle refrigeration can be realized by a Josephson-junction based qubit coupled to two thermal baths that are realized by two superconducting resonators with different frequencies ($f_h > f_c$) connected to a normal-metal resistor. Both resonators are coupled again to a metal resistor that has superconducting probes connected via NIS junctions. One experimental proposal is a Cooper-pair box qubit capacitively coupled to two resonators with different frequencies with a terminating resistor \cite{PhysRevApplied.17.064022}. Another proposal is to couple the qubit to the hot and cold baths through two different modes inside a cavity that couples inductively to a flux qubit. Figure~\ref{fig:1}(b,c) shows the schematic of the proposed device and the power measurements by NIS thermometers. The two modes ($f_1$ and $f_2$) are coupled independently to a resistor via a capacitor. The $f_1$ mode couples to a 'blue' resistor  and $f_2$ mode couples to a 'red' resistor. Powers dissipated in the two resistors due to cavity pumping are shown around $f_1 \sim 7.47~\mathrm{GHz}$ and $f_2 \sim 12.42~\mathrm{GHz}$ resonances. Blue lines indicate the power to the 'blue' resistor while the red lines indicate that to the 'red' resistor. 

\section{Conclusion}\label{sec13}

The key to the various demonstrated applications of SN interfaces has been in their relatively well understood physics, which allows one to tailor hybrids with desired properties in, e.g., devices for local probing and heat manipulation. This is especially true when using conventional superconductors and usual metallic conductors, where experimental reality meets the theoretical predictions often quantitatively. On top of that, integrating SN structures with modern circuit QED devices is relatively straightforward, opening possibilities for future experiments in this growing research area as well. It is also likely that the application areas widen towards higher temperatures from the convenient mK range thanks to advances in developing new superconducting materials at the nanoscale. Finally, superconductor-semiconductor hybrids share common features with SN hybrids, and the former ones form an active field of research that we did not cover in this article. This paper is by no means meant to be a balanced review of the field and therefore it does not cover all the work on SN interfaces either.

\bibliography{sn-bibliography}

\end{document}